\documentclass[a4paper,aps,prd,showpacs,preprintnumbers,twocolumn,nofootinbib]{revtex4}

\pdfoutput=1
\usepackage[normalem]{ulem} 
\usepackage{graphicx}
\usepackage{amsmath}
\usepackage{amsfonts}   
\usepackage{amssymb}  
\usepackage{mathrsfs}
\usepackage{epsfig,bbm}
\usepackage{bm}
\usepackage{color}
\usepackage{dsfont}
\begin{document}
\newcommand{\average}[1]{\langle{#1}\rangle_{{\cal D}}}
\newcommand{\dd}{{\rm d}}
\newcommand{\etal}{{\it et al.}}

\title{Cosmological evolution of the gravitational entropy of the large-scale structure}

\author{Giovanni Marozzi}
\email{giovanni.marozzi@unige.ch}
\affiliation{
		Universit\'e de Gen\`eve, D\'epartement de Physique Th\'eorique and CAP,
24 quai Ernest-Ansermet, CH-1211 Gen\`eve 4, Switzerland,}

\author{Jean-Philippe Uzan}
\email{uzan@iap.fr}
 \affiliation{
             Institut d'Astrophysique de Paris,
             Universit\'e Pierre~\&~Marie Curie - Paris VI,
             CNRS-UMR 7095, 98 bis, Bd Arago, 75014 Paris, France;\\
             Sorbonne Universit\'es, Institut Lagrange de Paris,
             98 bis, boulevard Arago, 75014 Paris, France,}

\author{Obinna Umeh}
\email{umeobinna@gmail.com}
 \affiliation{Department of Physics, University of the Western Cape,
Cape Town 7535, South Africa,}

\author{Chris Clarkson}
\email{Chris.Clarkson@uct.ac.za}
 \affiliation{Astrophysics, Cosmology \& Gravity Centre, and, Department of Mathematics \& \\
 Applied Mathematics, University of Cape Town, Cape Town 7701, South Africa.
 }

\begin{abstract}
We consider the entropy associated with the large-scale structure of the Universe in the linear regime,
where the Universe can be described by a perturbed Friedmann-Lema\^{\i}tre spacetime. In particular, we  compare  two different definitions proposed in the literature for the entropy using a spatial averaging prescription. 
For one definition, the entropy of the large-scale structure  for a given comoving volume always grows with time, both for a  CDM and a $\Lambda$CDM model. In particular, while it diverges for a CDM model, it saturates  to a constant value in the presence of a cosmological constant. The use of a light-cone averaging prescription in the context of the evaluation of the entropy is also discussed.
\end{abstract}

\maketitle

\section{Introduction}\label{sec1}

In the standard cosmological approach, the Universe is described by a homegeneous and isotropic solution of Einstein field equations, known as Friedmann-Lema\^{\i}tre (FL) spacetimes~\cite{pubook}. These solutions are expected to describe the Universe smoothed on cosmological scales. While this spacetime is easily identified in the early universe, matter clusters and structures grow under the effect of gravity so that the distribution of matter in our late time universe exhibit large inhomogeneities.

Over the past years significant activity has been devoted to the definition of averaging procedures~\cite{average1,average2,average3,GMNV} in order to construct a notion of a coarse-grained spacetime. Irrespective of the question of whether this procedure could explain the recent acceleration of the Universe, averaging methods are interesting since let us compare the evolution of a system described at different scales. In such a coarse-graining, information about the microscopic behaviour of the system is lost, which is at the origin of the notion of entropy. In essence, entropy  estimates the number of micro-states that correspond, after averaging, to a given macro-state.

The definition of gravitational entropy is still an open debate. While a suitable definition has been given in the context of the  thermodynamics of stationary  black holes~\cite{bh},  a well motivated and universally accepted  analogue has yet to be found in the cosmological context. With the evolution of the Universe, structure grows  and the Universe becomes more and more inhomogeneous. In order for the second law of thermodynamics to hold, the gravitational field itself shall carry entropy. It was argued in~\cite{Penrose1,Penrose2} that it has to be defined from the free gravitational field and thus be related to the Weyl tensor; more recently it was extended to a definition~\cite{CET} based on the Bel-Robinson tensor. This latter proposal reduces to the Bekenstein-Hawking entropy when integrated over the interior of a Schwarzschild black hole and increases as inhomogeneities grow.

While the concept of entropy arose from equilibrium thermodynamics, it has been also thought of as a measure of information. In terms of information theory the Kullback-Leibler divergence~\cite{KL}, for two probability distribution functions $p$ and $q$, is defined by
$$
D_{\rm KL}(p\vert q)\equiv\left<\ln\frac{p}{q} \right>_p
      =\int p(x)\ln\frac{p(x)}{q(x)}\dd x \,,
$$
and quantifies the amount of information lost when the data ($p$) is represented by the model ($q$). In cosmology, it was used in order to decide whether two cosmological models can be distinguished given a set of observational data~\cite{topology}.

Since the cosmological model of structure formation predicts the distribution of the density field (as a random variable), it has been proposed~\cite{entropyHBM}, in the context of averaging, to adapt the Kullback-Leibler divergence to a definition of relative information entropy that quantifies how the actual density field $\rho$ is different from its spatial average $\langle\rho\rangle_{\cal D}$ on a spatial domain ${\cal D}$ of proper volume $V_{\cal D}$, namely~\footnote{We add a factor $1/M_{Pl}$ with respect to the definition proposed in
\cite{entropyHBM} to obtain a dimensionless relative information entropy.}
\begin{equation}
\label{e.entropy1}
\frac{\cal S_{{\rm RI},\cal D}}{V_{\cal D}} \equiv\frac{1}{M_{Pl}}\left<\rho \ln\frac{\rho}{\langle\rho\rangle_{\cal D}} \right>_{\cal D}\,,
\end{equation}
with $M_{Pl}^{-2}=8 \pi G$.
It was conjectured that this function is increasing with cosmic time, which was checked for linear perturbations of a spatially Euclidean FL 
spacetime for a CDM model \cite{Li:2012qh} and for the comparison of a Lema\^{\i}tre-Tolman-Bondi (LTB) spacetime to its average~\cite{entropyHBM2,akerblom,ltb1,ltb2}. This is indeed a key property to consider Eq.~(\ref{e.entropy1}) as a valid definition of entropy.

This definition lies solely on the density field. 
In principle, we want to compare a spacetime $({\cal M},g)$ to  its average
$(\bar{\cal M},\bar g)$, and evaluate the quantity of information that has been lost.
The previous definition assumes that ${\cal M}$ can be foliated by a family of spatial {(or null)} hypersurfaces, which is related to the choice of the averaging procedure. It means that (i) the average depends on the choice of the slicing. The expression in Eq.~(\ref{e.entropy1}) involves an integral over quantities in the two spacetimes. It means that it implicitly involves a mapping between ${\cal M}$ and $\bar{\cal M}$ so that (ii) it may have a gauge dependence. 
{(iii) In principle, it also depends on the averaging procedure, whether it is spacelike~\cite{average1} or along the light-cone~\cite{GMNV}.} 
The expression (\ref{e.entropy1})  also depends solely on the matter distribution. After averaging, the matter distribution is indeed homogeneous but it may not be isotropic, as e.g. shown in Ref.~\cite{Marozzi:2012ib}, so that (iv) it may not capture the fact that the background spacetime may not be FL. Indeed, the geometry of the spacetime ${\cal M}$ is characterized  by its Riemann tensor $R_{\mu\nu\rho\sigma}$, that can be split as a Ricci contribution, $R_{\mu\nu}$, and a Weyl part $C_{\mu\nu\rho\sigma}$ (that can further be decomposed as an electric and magnetic parts, $E_{\mu\nu}$ and $B_{\mu\nu}$). The Ricci part is constrained by the matter distribution, via the Einstein field equations, but the knowledge of the density field alone does not allow one to reconstruct the Weyl part. As an example, consider ${\cal M}$ as a perturbed FL universe; it has non-vanishing $R_{\mu\nu}$, $E_{\mu\nu}$ and $B_{\mu\nu}$. After averaging $\langle\rho\rangle_{\cal D}$ is homogeneous. If $\bar{\cal M}$ is a FL universe then only $R_{\mu\nu}$ is non-zero while if $\bar{\cal M}$ is a Bianchi~I universe both $R_{\mu\nu}$ and $E_{\mu\nu}$  are non-zero, while they have the same $\langle\rho\rangle_{\cal D}$. It means that one part of the difference between the spacetimes is not included in the definition~(\ref{e.entropy1}). Part of this information is contained in the two scalars $C_{\mu\nu\rho\sigma}C^{\mu\nu\rho\sigma}$ and $C^*_{\mu\nu\rho\sigma}C^{\mu\nu\rho\sigma}$ (see Sec.~\ref{sec23} for definition) constructed from the Weyl tensor, which is indeed at the heart of the proposals~\cite{Penrose1,Penrose2,CET}. 
While not transparent in the definition (\ref{e.entropy1}), it can actually be shown (see Ref.~\cite{Li:2012qh}) that, for perturbations around an FL spacetime, this formula reduces in part to some combination of the Weyl scalars.

The paper is organized as follows. 
In Sec.~\ref{sec2} we define the foliation of our spacetime, 
the two averaging procedure we shall consider and introduce linear perturbation theory.
In Sec.~\ref{sec3} we present the different definitions of the gravitational entropy used, while in  Sec.~\ref{sec4} we investigate their time evolution. 
Finally, the results are discussed and compared in Sec.~\ref{sec5}.
Appendix~\ref{appA} presents the dynamics of the background spacetime, and Appendix~\ref{appB} summarizes the definition of gauge invariant degrees of freedom.

\section{Averaging procedures}\label{sec2}

Averaging procedures rely on a choice of observers/foliation of spacetime, described in Sec.~\ref{sec21}. In the following, we consider two procedures based either on spatial sections or on null sections (Sec.~\ref{sec22}).

\subsection{Spacetime foliation}\label{sec21}

To define our formalism let us introduce a 1+3 splitting of the Universe~\cite{1plus3} associated with a general reference timelike 
congruence $n^\mu$ that defines a class of observers and the relative foliation of spacetime. 
The 3-dimensional spacelike hypersurfaces normal to $n^\mu$ can then be defined by the equation $S({\bm x}, t)-S_0=0$, with $S({\bm x}, t)$ a scalar field and $S_0$ a constant. Then
\begin{equation}
\label{DefnA}
n_{\mu} \equiv -  \frac{\partial_{\mu} S}{ (-
 \partial_{\rho} S \partial_{\nu} S ~ g^{\rho\nu}) ^{1/2}},
\end{equation} 
is normalized as $n_{\mu} n^{\mu} = -1$. This allows us to define $h_{\mu\nu}$, the projector on these hypersurfaces, as
\begin{equation}
  h_{\mu\nu} = g_{\mu\nu} + n_{\mu}n_{\nu}
\end{equation}
which satisfies by construction $h_{\mu\rho}h^{\rho}_{\nu} = h_{\mu\nu}$ and $h_{\mu\nu}n^{\mu} = 0$. 
Furthermore, one can define 
the expansion $\Theta$, shear $\sigma_{\mu\nu}$ and vorticity $\omega_{\mu\nu}$ of the flow as
\begin{eqnarray}
   \Theta_{\mu\nu} &\equiv& h^\alpha_\mu h^\beta_\nu \nabla_\alpha n_\beta\\
                               & =&\frac{1}{3} h_{\mu\nu}\Theta+\sigma_{\mu\nu}+\omega_{\mu\nu} .
\end{eqnarray}
They are explicitly given by
\begin{equation}\label{e.theta}
    \Theta \equiv \nabla_\mu n^\mu,
\end{equation}    
\begin{equation}
  \sigma_{\mu\nu}\equiv h^\alpha_\mu h^\beta_\nu \left[\nabla_{(\alpha} n_{\beta)}-
  \frac{1}{3} h_{\alpha\beta}
\nabla_\tau n^\tau \right],
\end{equation}
\begin{equation}
\omega_{\mu\nu}\equiv h^\alpha_\mu h^\beta_\nu \nabla_{[\alpha} n_{\beta]}\,.
\end{equation}
Indeed, the assumption of Eq. (\ref{DefnA}) implies that the vorticity strictly vanishes, $\omega_{\mu\nu}=0$. 

In practice, perturbations grow significantly only during the matter-dominated era, so that one can restrict the analysis to dust-filled universes, eventually with a cosmological constant. In such a situation one can pick up a foliation defined by a congruence $n_\mu$ corresponding to the four-velocity of a geodesic observer, which accidentally coincides with the four-velocity $u_\mu$ of comoving observers, i.e. $n_\mu=u_\mu$.\footnote{In general $n_\mu$, which defines a general reference flow, and $u_\mu$, which defines the four-velocity of the observers comoving with the matter, may be different (see Ref.~\cite{GMV2} for details).}

The shear tensor can be then expressed as
\begin{equation}
 \sigma_{\mu\nu}=\Theta_{\mu\nu}-\frac{1}{3} h_{\mu\nu}\Theta \,,
\end{equation}
and it follows that the scalar shear takes the form
\begin{equation}
  \sigma^2\equiv\frac{1}{2} \sigma^\mu_\nu \sigma^\nu_\mu=
\frac{1}{2} \left(\Theta^\mu_\nu \Theta^\nu_\mu-\frac{1}{3}
\Theta^2\right)\, .
\label{sigma2}
\end{equation}

\subsection{Spatial and light-cone averaging}\label{sec22}

On one hand, we consider a spatial averaging procedure~\cite{average1}, entirely  based on a slicing of spacetime by spatial hypersurfaces. The spatial average of any scalar quantity $A$ on a domain ${\cal D}$ is then defined as
\begin{equation}
 \average{A(\eta,{\bm x})} = \frac{1}{V_{\cal D}}\int_{\cal D} \sqrt{|h|}\, A(\eta,{\bm x})\dd^3{\bm x} \,,
\label{averagepb}
\end{equation}
where $V_{\cal D}$ is the volume of the domain, defined by the requirement that $\average{1}=1$, 
and $h$ is the determinant of the induced metric $h_{\mu\nu}$ on the averaging hypersurface. 
Such a spatial average is associated with the general reference timelike congruence $n^\mu$
of Eq.~(\ref{DefnA}) if and only if the average is performed in the gauge where $S(\eta, {\bm x})$ is 
homogeneous (see Refs.~\cite{GMV1,GMV2,Marozzi}). 

On the other hand, cosmological observations are usually restricted on the past light-cone, since most of the relevant signals are of electromagnetic origin. Hence, when we look to cosmological observables, the averaging procedure should be possibly referred to a null hypersurface coinciding with our past light-cone or to the null surface obtained from  the intersection of our past light-cone with some fixed-time spacelike hypersurface. Let us consider this latter possibility. Following Ref.~\cite{GMNV} we obtain that the averaging of any scalar $A(\eta, {\bm x})$ over the 2-sphere embedded in our past light-cone, defined by a null scalar $V(\eta, {\bm x})$ (i.e., such that $\partial_\mu V \partial^\mu V=0$) equal to a constant, and corresponding to its intersection with the spacelike hypersurface   $S(\eta, {\bm x})=S_0$, is given by
\begin{eqnarray}
\langle A(\eta,{\bm x}) \rangle_{V_0,S_0}&=& \frac{1}{V_{\cal S}} \int_{{\cal M}} \sqrt{-g}~ \delta (V_0-V)  \delta(S-S_0)  \nonumber\\
 &&\qquad A(\eta,{\bm x})\, |\partial_\mu V \partial^\mu S| \,\dd^4x \,, 
\label{averagepb-LC}
\end{eqnarray}
where ${\cal M}$ is the 4-dimensional spacetime,  $V_{\cal S}$ is the volume of the 2-sphere embedded in the light-cone, defined by the requirement that $\langle 1 \rangle_{V_0,S_0}=1$, and $g$ is the determinant of the four dimensional metric $g_{\mu\nu}$.

\subsection{Linear peturbation theory}\label{sec23}

The standard scalar-vector-tensor  decomposition \cite{pubook} of a perturbed FL spacetime have metric components
\begin{eqnarray} 
   \delta^{(1)}  g_{00} &=& -2 a^2 \alpha\nonumber\\
   \delta^{(1)}  g_{i0}&=& -\frac{a^2}{2} B_i=-\frac{a^2}{2} (\partial_i \beta+\bar{B}_i) \,,\nonumber\\
\delta^{(1)}g_{ij} &=& a^2 \left[ -2 \psi \delta_{ij} + D_{i j} E +
\partial_{(i}\bar{\chi}_{j)} +\frac{1}{2} \bar{h}_{i j} \right] \!,
\label{General_metric}
\end{eqnarray}
with $D_{ij}\equiv \partial_i \partial_j-\frac13\delta_{ij}\Delta$. 
We then have 4 scalar degrees of freedom ($\alpha$, $\beta$, $\psi$ and $E$), 2 transverse vectors ($\bar{B}_i$  and $\bar{\chi}_i$ with $\partial^i \bar{B}_i=0$,  $\partial^i \bar{\chi}_i=0$) with 4 degrees of freedom, and a traceless and transverse tensor ($\bar{h}_{ij}$ with $ \partial^i \bar{h}_{ij}=0=\bar{h}_i^i$) with 2 degrees of freedom.\footnote{The definition of gauge invariant degrees of freedom 
are summarized in Appendix~\ref{appB}.}

Let us stress that first order perturbation theory is sufficient to obtain the general expression for the shear in Eq.~(\ref{sigma2}) up to second order. 
Since second order perturbations contribute only to third or fourth order to $\sigma^2$ (see Appendix~\ref{appB}).

In the following we shall use the synchronous gauge and neglect vector and tensor perturbations. We then have
\begin{equation}
\label{e.synchronous}
 \dd s^2= a^2\left\{-\dd\eta^2 + \left[(1-2\psi)\delta_{ij}+D_{ij} E\right]
 \dd x^i\dd x^j \right\}.
\end{equation}
It is clear from Eqs.~(\ref{b1}-\ref{b4}) that we can then write the Bardeen potentials $\Psi$ and $\Phi$ as
\begin{equation}
 \Psi =\psi+\frac{1}{6} \Delta E +\frac{{\cal H}}{2} E', \qquad
 \Phi =-\frac{{\cal H}}{2} E'- \frac{E''}{2} \,,
\end{equation}
where the prime denotes the derivative with respect to conformal time and ${\cal H}=a'/a$.

Let us now introduce the Weyl tensor defined as 
\begin{eqnarray}
C_{\mu\nu\lambda\rho}&=&R_{\mu\nu\lambda\rho}+\frac{1}{2}\left(g_{\mu\rho}R_{\nu\lambda}+g_{\nu\lambda}R_{\mu\rho}-g_{\mu\lambda}R_{\nu\rho} \right. \nonumber \\
& & \left.
-g_{\nu\rho}R_{\mu\lambda}\right)+\frac{1}{6}\left(g_{\mu\lambda}g_{\nu\rho}-g_{\mu\rho}g_{\nu\lambda}\right) R \,,
\end{eqnarray}
where $R_{\mu\nu\lambda\rho}$ and $R_{\mu\nu}$ are the Riemann and Ricci tensors, while $R$ is the Ricci scalar.
We can then define the dual of the Weyl tensor as  $C^*_{\alpha\mu\nu\beta}=\frac{1}{2} \eta_{\alpha\mu\tau\gamma}C^{\tau\gamma}_{\quad\nu\beta}$, where $\eta_{\alpha\mu\tau\gamma}=\sqrt{-g} \epsilon_{\alpha\mu\tau\gamma}$ is the four dimensional volume element.

In terms of the Bardeen potentials, in any gauge and for vanishing anisotropic stress, the Weyl scalar $C_{\mu\nu\lambda\rho}C^{\mu\nu\lambda\rho}$ takes the simple form
\begin{equation}\label{WTc}
  C_{\mu\nu\lambda\rho}C^{\mu\nu\lambda\rho}=\frac{8}{a^4} D_{ij} \Phi D^{ij} \Phi\,.
\end{equation}

Let us relate the contraction of the Weyl tensor in Eq.~(\ref{WTc}) to the shear of a free-falling observer given in the synchronous 
gauge by
\begin{equation} 
\sigma^2=\frac{1}{8 a^2} D_{ij} {E}' D^{ij} {E}'\,,
\label{sigmaSG}
\end{equation}
which shows that $\sigma^2$ and $C_{\mu\nu\lambda\rho}C^{\mu\nu\lambda\rho}$ are not independent quantities.
If we consider a general $\Lambda$CDM model we have
 \begin{eqnarray}
\psi(\eta,\vec{x})&=&\frac{2}{9 \mathcal{H}^2 \Omega_m} \nabla^2 \Phi(\eta, \vec{x})+\frac{5}{3} \Phi(\eta_{in}, \vec{x}),\\
E(\eta, \vec{x})&=& -\frac{4}{3 \mathcal{H}^2 \Omega_m} \Phi(\eta, \vec{x})\,.\label{chiGeneral}
\end{eqnarray}
Using Eq.~(\ref{chiGeneral}) and the background dynamics (see Appendix A), we obtain the following relation which connects the shear to the gravitational potential
\begin{eqnarray} 
\label{sigmaSGfunctionGP}
\sigma^2&=&\frac{2}{9} \left(\frac{1}{a_0 {\cal H}_0^2 \Omega_{m0}}\right)^2 \left[{\cal H} D_{ij} \Phi+D_{ij} \Phi'\right] 
 \nonumber \\
& &\quad \times \left[{\cal H} D^{ij} \Phi 
+D^{ij} \Phi'\right]\,.
\end{eqnarray}
The relation between the Weyl tensor in Eq.~(\ref{WTc}) and the shear of a free-falling observer is then given by 
\begin{eqnarray}
& & \!\!\!\!\!\! \!\!\!\!\!\!C_{\mu\nu\lambda\rho} C^{\mu\nu\lambda\rho}=\frac{8}{a^4} \left[\frac{2}{9} \left(\frac{1}{a_0 {\cal H}_0^2 \Omega_{m0}}\right)^2 
{\cal H}^2\right]^{-1}
\sigma^2 \nonumber \\
& &\,\,\,\,\,\,
-\frac{16}{a^4 {\cal H}}D_{ij} {\Phi} D^{ij} \Phi'
-\frac{8}{a^4 {\cal H}^2} D_{ij} \Phi' D^{ij} \Phi'\,.
\label{GeneralizationCvsShear}
\end{eqnarray}
In a CDM model, the gravitational potential is constant, so that the contraction of the Weyl tensor and the shear are related simply by a time dependent factor. In a $\Lambda$CDM model, one needs to include the terms arising from the decay of the gravitational potential.

\section{Definitions of the entropy}\label{sec3}

\subsection{Definition from the density field}\label{sec3-DD}

As discussed in the introduction, the first idea to define a relative entropy between two spacetimes~\cite{entropyHBM} followed the definition of the  Kullback-Leibler divergence~\cite{KL} in information theory. It allows one to quantify whether two density fields $\rho$ and  $\langle\rho\rangle_{\cal D}$ can be distinguished and is defined\footnote{In the following we consider the spatial averaging prescription $<....>_{\cal D}$ in our definition, we will show in Sec. IV that this is indeed the right averaging prescription for the evaluation of the entropy.} by Eq.~(\ref{e.entropy1}).

When working in perturbation the density field can be expanded as $\rho = \rho^{(0)}+\rho^{(1)}+\rho^{(2)}$, so that at leading order \cite{Li:2012qh}
\begin{eqnarray}
\label{pS}
   \frac{\cal S_{{\rm RI},\cal D}}{V_{\cal D}}&=&\frac{1}{2 M_{Pl}}\frac{\langle(\rho^{(1)})^2\rangle_{\cal D}-\langle\rho^{(1)}\rangle^2_{\cal D}}{\rho^{(0)}} \,.
\end{eqnarray}
As can be seen from this equation, 
the relative information entropy can be obtained at leading order using only first order perturbation theory. In particular, this is given by the 
variance of the energy density.

Eq.~(\ref{pS}) is valid independently of the matter content of the Universe. Following our introductory considerations, the density field that constraints the Ricci part of the Riemann tensor is the total energy density. Therefore, to define the entropy we shall use the total density energy of the Universe.

Let us now evaluate Eq.~(\ref{pS}) in a $\Lambda$CDM universe.  In the synchronous gauge, the first order perturbation of the energy density is given by the Poisson equation (this corresponds to the matter perturbation because the cosmological constant cannot be perturbed by definition)
\begin{equation} 
  \rho^{(1)}(\eta,\vec{x})=\frac{2}{a^2} M_{Pl}^2 \nabla^2 \Phi(\eta,\vec{x}) \,.
  \end{equation}
As a consequence Eq.~(\ref{pS}) becomes
\begin{equation}
\frac{\cal S_{{\rm RI},\cal D}}{V_{\cal D}}=\frac{2}{3} \frac{M_{Pl}}{{\cal H}^2 a^2} \left[\langle\left(\nabla^2 \Phi\right)^2\rangle_{\cal D}-
\langle \nabla^2 \Phi \rangle_{\cal D}^2\right]\,.
\label{SRIfirst}
\end{equation}
Let us now use the expression of Eq.~(\ref{WTc}) for the Weyl tensor in a perturbed FL metric to rewrite Eq.~(\ref{SRIfirst}) in a useful form.
After some simple algebraic manipulations we obtain that
\begin{eqnarray}
& &\!\!\!\!\!\!\!\frac{\cal S_{{\rm RI},\cal D}}{V_{\cal D}}=\frac{9}{4} M_{Pl} \left[ \frac{a^2}{18 {\cal H}^2 } \langle 
C_{\mu\nu\lambda\rho}C^{\mu\nu\lambda\rho} \rangle_{\cal D} +\frac{4}{9 a^2 {\cal H}^2}
 \right. \nonumber \\
& & \left. \times \left(\langle\left(\nabla^2 \Phi\right)^2\rangle_{\cal D}-\langle \partial_i\partial_j \Phi \partial^i\partial^j \Phi \rangle_{\cal D}
-\frac{2}{3}
\langle \nabla^2 \Phi \rangle_{\cal D}^2\right)\right]\,. \nonumber \\
& & 
\label{GeneralSoverVrhot}
\end{eqnarray}

As shown in Ref.~\cite{Li:2012qh} the term in the second line of Eq.~(\ref{GeneralSoverVrhot}) is related to the so-called  kinematical backreaction $ {\cal Q}_{\cal D}$ (see Ref.~\cite{average1}), given by
\begin{equation}\label{e.QD}
 {\cal Q}_{\cal D} \equiv \frac{2}{3}\left(\average{\Theta^2}-\average{\Theta}^2 \right) - 2\average{\sigma^2} \,,
\end{equation}
in the CDM case.
Therefore, for a free falling observer in the synchronous gauge and considering a CDM model we can rewrite Eq.~(\ref{GeneralSoverVrhot}) in the following way \cite{Li:2012qh} 
\begin{equation}
\label{GeneralSoverVrhotCDM}
\frac{\cal S_{{\rm RI},\cal D}}{V_{\cal D}}=\frac{9}{4} M_{Pl} \left[ \frac{a^2}{18 {\cal H}^2 }  \langle 
C_{\mu\nu\lambda\rho}C^{\mu\nu\lambda\rho} \rangle_{\cal D}+
 {\cal Q}_ {\cal D}\right]\,.
\end{equation}

The entropy~(\ref{GeneralSoverVrhot}) is the average of a combination of scalar quantities and it is gauge invariant under a gauge transformation (see Ref.~\cite{Marozzi} for the possible gauge dependence coming from the averaging prescription). Indeed, this scalar combination is zero at zero and first orders (see Eq. (\ref{pS})), and therefore gauge invariant at leading order under a gauge transformation.

\subsection{Definition from the Bell-Robinson tensor}\label{sec3-BR}

Ref.~\cite{CET} suggested a thermodynamically motivated measure of the gravitational entropy based on the Bel-Robinson tensor,
\begin{equation}
T_{\mu\nu\rho\sigma}=\frac{1}{4}\left(C_{\alpha\mu\nu\beta}{{C^\alpha}_{\rho\sigma}}^\beta+
C^*_{\alpha\mu\nu\beta}{{C^{*\alpha}}_{\rho\sigma}}^\beta
\right).
\end{equation}
A measure of gravitational entropy constructed from this tensor was also considered in \cite{PL,PC}, using an integral over conformal time of the super-energy density $W$ defined by 
\begin{equation}
W=T_{\mu\nu\rho\sigma}n^\mu n^\nu n^\rho n^\sigma \,.
\end{equation}
Note that this super-energy density $W$  is observer dependent and non-negative - which is not a problem per se, since the entropy is also observer dependent.

Following Ref.~\cite{CET} and imposing the following five conditions for the entropy: non-negative, vanishing only if $C_{\mu\nu\rho\sigma}=0$, it should measure the local anisotropy of the free gravitational field, reproduce the Bekenstein-Hawking entropy of a black hole and increase monotonically as structure forms, one can define a  thermodynamically motivated measure of the gravitational entropy.  For the case of a perturbed FL spacetime with Euclidean spatial sections, this takes the form~\cite{CET}
\begin{equation}
S_{G,D}'=4 \pi M_{Pl}^2 \lambda \frac{a}{{\cal H}} \int_D \frac{d}{d\eta}\left(a^3 \sqrt{\frac{W}{6}}\right)d^3{\bf x} ,
\label{DerSCliftonetall-first}
\end{equation}
with $\lambda$ a constant, and where we integrate over a comoving volume ${\cal V}_{\cal D}$.
Indeed, this differs from the definition~(\ref{e.entropy1}) of the relative entropy between two spacetimes. 
In particular, it seems to depend on the whole history of the spacetime, being present on the left hand side of 
Eq.(\ref{DerSCliftonetall-first}) a time derivative. 
But, since the entropy of a spacetime with vanishing Weyl tensor is zero, the entropy of a FL spacetime vanishes, so that it can also be considered as the relative entropy with respect to the background FL spacetime.

We now want to compare this result with our previous result for the case of a freely falling observer in a background spacetime plus first order perturbations. In this particular case, one easily concludes that 
\begin{equation}
W=\frac{1}{4}C_{\mu 0 0 \rho}C^{\mu 0 0 \rho} = \frac{1}{32}
C_{\mu\nu\lambda\rho}C^{\mu\nu\lambda\rho} 
\end{equation}
namely the super energy density is equal (up to a constant) to the Weyl scalar, having the part that comes from the dual of the Weyl tensor zero contribution. As a consequence the gravitational entropy of Ref.~\cite{CET} reduces to 
\begin{equation}
S_{G,D}'=4 \pi M_{Pl}^2 \lambda \frac{a}{{\cal H}} \int_D \frac{d}{d\eta}\left(a^3 \sqrt{\frac{C_{\mu\nu\lambda\rho}C^{\mu\nu\lambda\rho} }{192}}\right)d^3{\bf x} .
\label{DerSCliftonetall-second}
\end{equation}

\section{Time evolution of the entropy}\label{sec4}

We can now compare the two definitions of entropy of Eqs. (\ref{GeneralSoverVrhot}) and (\ref{DerSCliftonetall-second}), obtained in Sec.~\ref{sec3-DD} and Sec.~\ref{sec3-BR}, to describe  the large-scale structure of the Universe, in a $\Lambda$CDM model.

In the description adopted here, the perturbations are stochastic fields, usually with initially Gaussian  statistics. It follows that, for example, the spatially averaged quantities are also stochastic fields. If $X$ is a function of the perturbations and $\langle X \rangle$ its average 
on a given domain then, from a theoretical point of view, we only have access to the distribution of $\langle X \rangle$, that is to $\overline{\langle X \rangle}$, which is the ensemble average of  $\langle X \rangle$.  Hence, one has to perform the ensemble average of the quantities defined in Secs.~\ref{sec3-DD} and~\ref{sec3-BR}.

To compare the definitions of entropy, we have now to choose an averaging procedure. In general, one has two possibilities. The first is to average over a volume embedded in a spatial hypersurface, for example the one of constant redshift $z$, while the second is to average over a two-sphere defined as the intersection of our past or future light-cone with this spatial hypersurface. These two averaging procedures turns out to be equivalent for terms of the kind $\overline{\langle (f(\Phi))^2 \rangle}$, with $f(\Phi)$ a linear function of the gravitational potential.  This is due to the fact that in this case we average a quantity already of the second order in perturbation theory (and the averages can be performed only at the background level, if one stops at second order), for a case in which the shape of the domain of integration is not important (see Eq.~(5.3) of Ref.~\cite{BenDayan:2012pp} for the light-cone averaging case and Ref.~\cite{Marozzi:2012ib} for the spatial averaging case).  On the other hand, for terms of the kind $\overline{\langle f(\Phi)\rangle^2}$ the average depends on the shape of the domain of integration and the two procedures give different results (see Eq.~(5.4) of Ref.~\cite{BenDayan:2012pp} for the light-cone averaging case and Ref.~\cite{Clarkson:2009hr} for the spatial averaging case). Therefore, it is important to specify which prescription has to be used to obtain a physically meaningful result. As a guideline we consider the fact that our entropy should describe the entropy of the large-scale structure of the Universe; that is, it should  characterize the Universe as a whole and should be averaged over an extended region. One can then immediately exclude the light-cone averaging prescription for several reasons. First, the light-cone averaging prescription corresponds to an average over a two-sphere with dimensions dependent from the value of redshift considered, after we fix the observer. For example, for a redshift equal to $z=0.01$ the region of integration (the two-sphere) would be all inside our local universe, where non-linearities became extremely important. Even worse,  the region of integration goes to zero at redshift equal to zero.
Furthermore, considering different times of observation for a given observer (i.e. different light-cones) the entropy at a given moment would be evaluated using different regions of integration.  

As a consequence, it is easy to understand why the light-cone averaging prescription cannot be applied in this context. The right choice for the case in consideration is then the spatial averaging prescription. 
This should not surprise us because the entropy is not an observable in the standard way~\footnote{See Ref.~\cite{ApLC} for the case of cosmological observables, such as the second order luminosity 
distance/redshift relation~\cite{dL2ord}, where  the application of the light-cone averaging prescription is necessary.}.

For both  definitions of the entropy, obtained in Secs.~\ref{sec3-DD} and~\ref{sec3-BR}, the key quantity to be evaluated is 
$\overline{\langle C_{\mu\nu\lambda\rho}C^{\mu\nu\lambda\rho} \rangle_{\cal D}}$. Now, let us consider a peculiar-gravitational potential described by a linear power spectrum of fluctuations, with the transfer function given in  Ref.~\cite{EH} (without the contribution from the baryons) and the cosmological parameters from PLANCK~\cite{PLANCK}. Considering the Fourier expansion of the first-order gravitational potential we obtain, in Fourier space (and independently from the integration domain ${\cal D}$; see Ref.~\cite{Marozzi:2012ib})
\begin{equation}
\overline{\langle C_{\mu\nu\lambda\rho}C^{\mu\nu\lambda\rho} \rangle_{\cal D}}=\frac{16}{3 a^4} \int \frac{d k}{k} k^4 \mathcal{P}_\varphi(\eta, k)\,,
\label{Cscalarmodek}
\end{equation}
with $\mathcal{P}_\varphi(\eta, k)=\frac{k^3}{2 \pi^2} |\varphi_k|^2$ the power spectrum of the gravitational potential. One can then easily see that the integration in Eq.~(\ref{Cscalarmodek}) has an ultraviolet divergence if one takes the linear power spectrum defined in  Ref.~\cite{EH}. We shall thus consider this as an effective description and hereafter we will assume a cut-off $k_{\rm UV}=0.1 h {\rm Mpc}^{-1}$ to stay within the linear regime and regularize the ultraviolet divergence.

To assume a constant cut-off $k_{UV}$ is enough for our purpose. In fact, despite in general the non-linearity scale evolve 
in time, we are mainly interested in a $\Lambda$CDM model, starting from when the cosmological constant begins to dominate. For this case the 
non-linearity scale stays nearly constant and our assumption is justified. As we will see in the following, this is the case where the 
different entropy proposals here considered behaviour in a different way one with respect to the other.

While the result in Eq.~(\ref{Cscalarmodek}) is enough to evaluate the gravitational entropy of Eq.~(\ref{DerSCliftonetall-second}), for the relative information entropy of Eq.~(\ref{GeneralSoverVrhot}) we have to evaluate also the terms present in the second line. In Fourier space the terms $\overline{\langle\left(\nabla^2 \Phi\right)^2\rangle_{\cal D}}$ and $\overline{\langle \partial_i\partial_j \Phi \partial^i\partial^j \Phi \rangle_{\cal D}}$ cancel each other, while the third term $\overline{\langle \nabla^2 \Phi \rangle_{\cal D}^2}$ gives a result dependent from the window function used \cite{Clarkson:2009hr}.
If we consider a top-hat window function of radius $R$ to smooth the field, i.e. we have
$$
W_R(|{\bf x}|)=\left(\frac{4}{3} \pi R^3\right)^{-1} \Theta(R-|{\bf x}|) \,,
$$
with $\Theta$ the Heavyside function, we then have (see, for example, Ref.~\cite{Marozzi:2012ib})
\begin{equation}
\overline{\langle \nabla^2 \Phi \rangle_{\cal D}^2}=  \int \frac{d k}{k} k^4 \mathcal{P}_\varphi(\eta, k)  \left(3 \frac{j_1(k R)}{kR}\right)\,,
\label{ResTer}
\end{equation}
with $j_1(x)=1/x^2 (\sin(x)-x \cos(x))$ a spherical Bessel function. As can be shown, comparing Eq. (\ref{ResTer}) with respect to 
Eq. (\ref{Cscalarmodek}),  the contribution of $\overline{\langle \nabla^2 \Phi \rangle_{\cal D}^2}$ to the total value of the relative information entropy is negligible as soon as we consider a window function with, for example, a radius $R$ at least one order of magnitude larger than the cut-off scale $k_{UV}$. This is a well motived physical choice for the case under consideration because we want to stay inside the linear regime and the ultraviolet cut-off chosen is, indeed, the present threshold of the linear regime. Therefore, we will neglect  this term hereafter in the evaluation of the information entropy of  Eq.~(\ref{GeneralSoverVrhot}).
\vskip0.25cm

\subsection{Relative Information Entropy}

Let us begin with some general considerations. Considering Eq.~(\ref{GeneralSoverVrhot}), we have that $S_{{\rm RI},{\cal D}}/V_{\cal D}$ and $\langle C_{\mu\nu\lambda\rho}C^{\mu\nu\lambda\rho} \rangle_{\cal D}$ are proportional only if the terms in the second line of Eq.~(\ref{GeneralSoverVrhot}) give negligible contribution. As showed in the previous section, this is the case for the case of a Universe described by a perturbed FL spacetime when we average over a large window function. 

As a consequence, using Eq.~(\ref{GeneralSoverVrhot}), we can conclude that
\begin{equation}
\frac{S_{{\rm RI},{\cal D}}}{{\cal V}_{\cal D}} \simeq M_{Pl} \frac{a^5}{8  {\cal H}^2} \overline{\langle C_{\mu\nu\lambda\rho}C^{\mu\nu\lambda\rho} \rangle_D}\label{Volume_Entropy_Lambda}
\end{equation}
is a good approximation. Then, using the result of Eq.~(\ref{Cscalarmodek}), we determine the behaviour of the relative information entropy per unit comoving volume ${\cal V}_{\cal D}$. It is depicted on Fig.~\ref{Fig1} for a CDM model (left panel) and a $\Lambda$CDM model (right panel). As shown by these two figures, the expression of Eq.~(\ref{Volume_Entropy_Lambda}) is monotonically increasing with the time only for a CDM model, i.e. as long as the cosmological constant is vanishing. The fact that the relative information entropy is not a valid definition of entropy for a $\Lambda$CDM universe can be understood by the fact that in the limit for which the proper time goes to infinity (equivalent to $z \rightarrow -1$) the contribution of the cosmological constant is more and more dominant. The Universe, both at the background level and at the perturbed level, is attracted toward a de Sitter spacetime. Therefore, the relative information entropy between these two spacetimes stops growing and then decreases asymptotically to zero because all scalar perturbations are washed out. The turn-over corresponds roughly at the time when the cosmological constant starts dominating the cosmic expansion.
  
To conclude, it is easy to see from Eqs.~(\ref{Volume_Entropy_Lambda}) and~(\ref{Cscalarmodek}), that the relative information entropy evolves as $\sim a^2(\eta)$ for the CDM case, as already stated in Ref.~\cite{Li:2012qh}.

Let us stress that the shape of the relative information entropy in Fig.~\ref{Fig1}, and of the gravitational entropy in the figures in the follow,  are independent from the value of the ultraviolet cut-off used. This is a consequence of the fact that we use a power-spectrum with a linear transfer function (as given in Ref.~\cite{EH}) for which the $k$ and time dependence are factorized. If we had considered a non-linear transfer function (like the ones in Refs.~\cite{Smith:2002dz,Takahashi}) then the $k$ and time dependence cannot be factorized anymore, and the shape of our figures would be slightly dependent from the value of the choice of the ultraviolet cut-off.

\newpage

\begin{widetext}
\hspace{1mm}
\begin{figure}[h!]
\centering
\includegraphics[width=8cm]{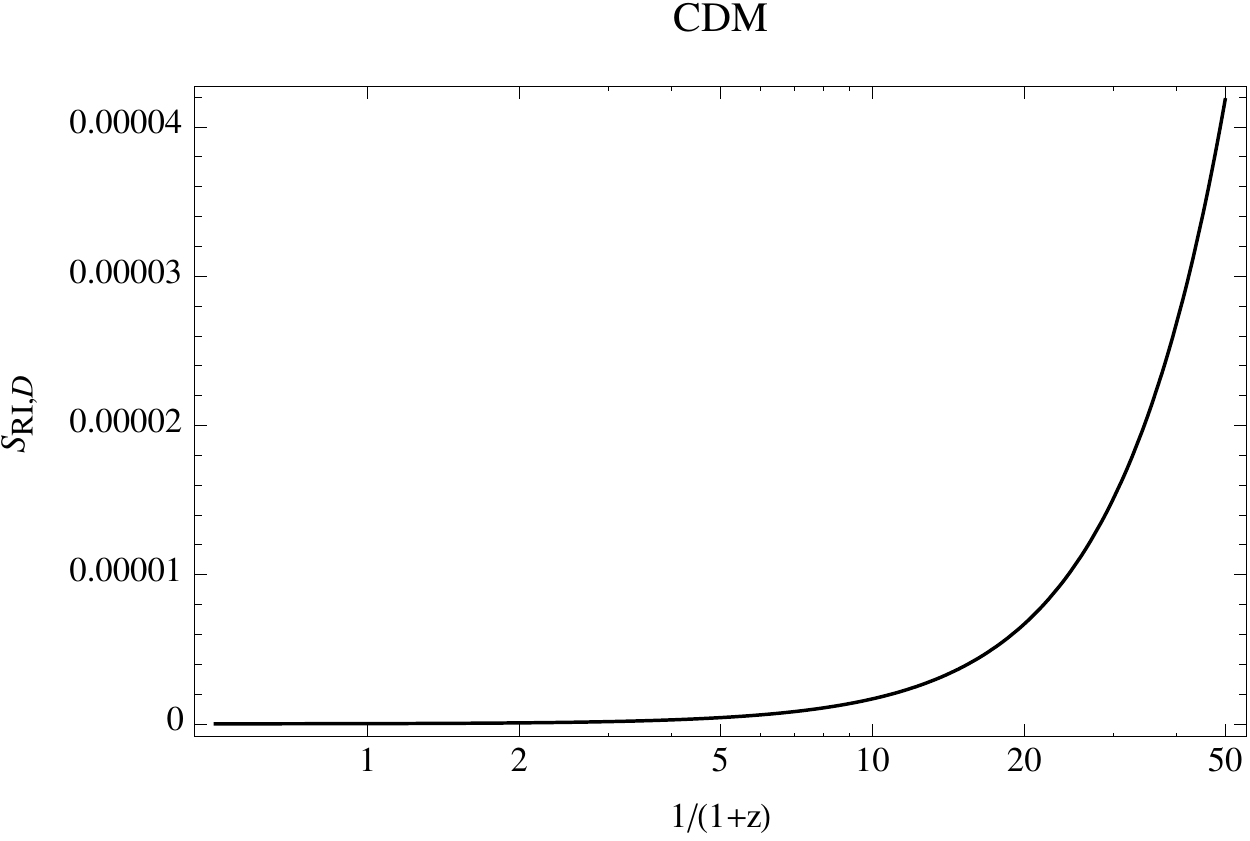}~~~~~~~~~
\includegraphics[width=8cm]{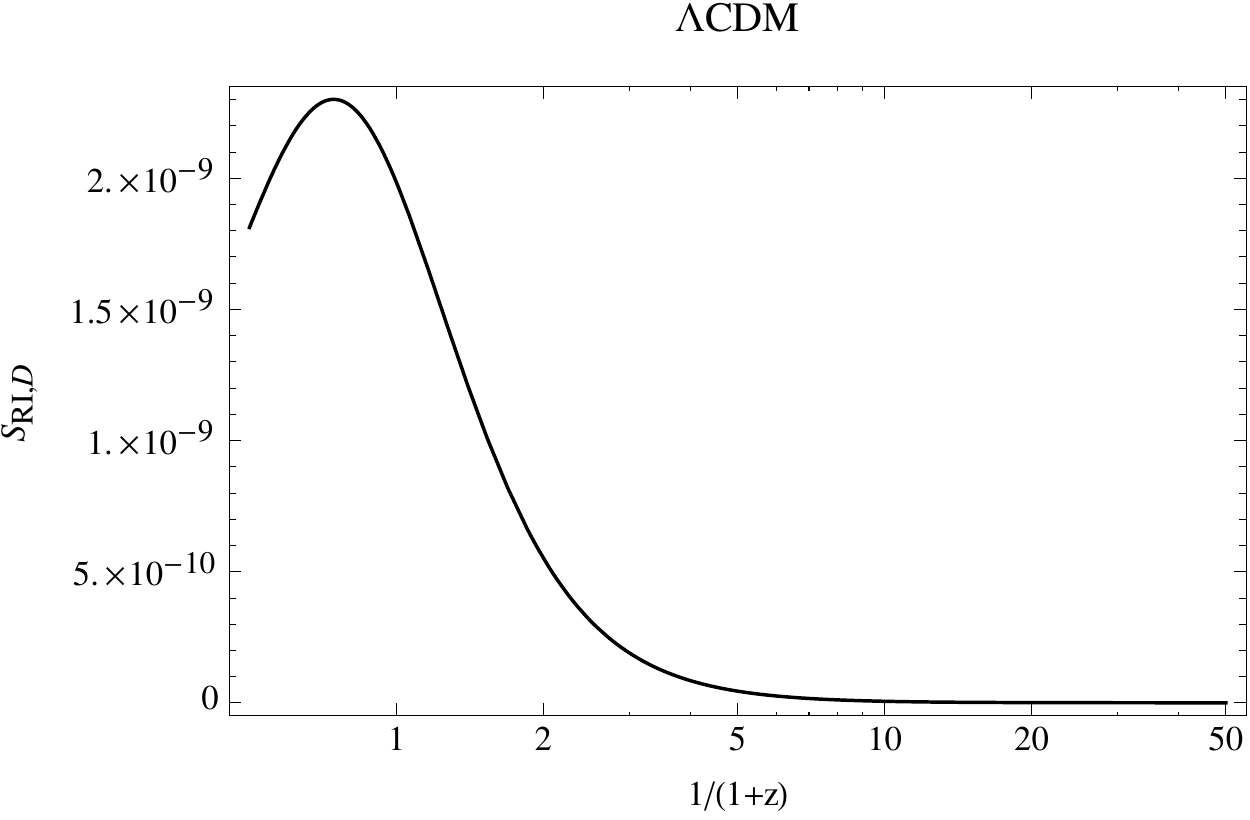}
\caption{Evolution of the relative information entropy per unit comoving volume, obtained from  Eqs.~(\ref{Volume_Entropy_Lambda}),  for the CDM model (left panel) and a $\Lambda$CDM model (right panel), setting $8 \pi G=1$ and assuming an ultraviolet cut-off $k_{\rm UV}=0.1 h {\rm Mpc}^{-1}$, as a  function of $1/(1+z)$.}\label{Fig1}
\end{figure}
\end{widetext}

\subsection{Gravitational Entropy}

Let us now perform a similar analysis for the gravitational entropy defined in Ref.~\cite{CET}, and described in Sec.~\ref{sec3-BR}.

\begin{figure}[h!]
\centering
\includegraphics[width=8cm]{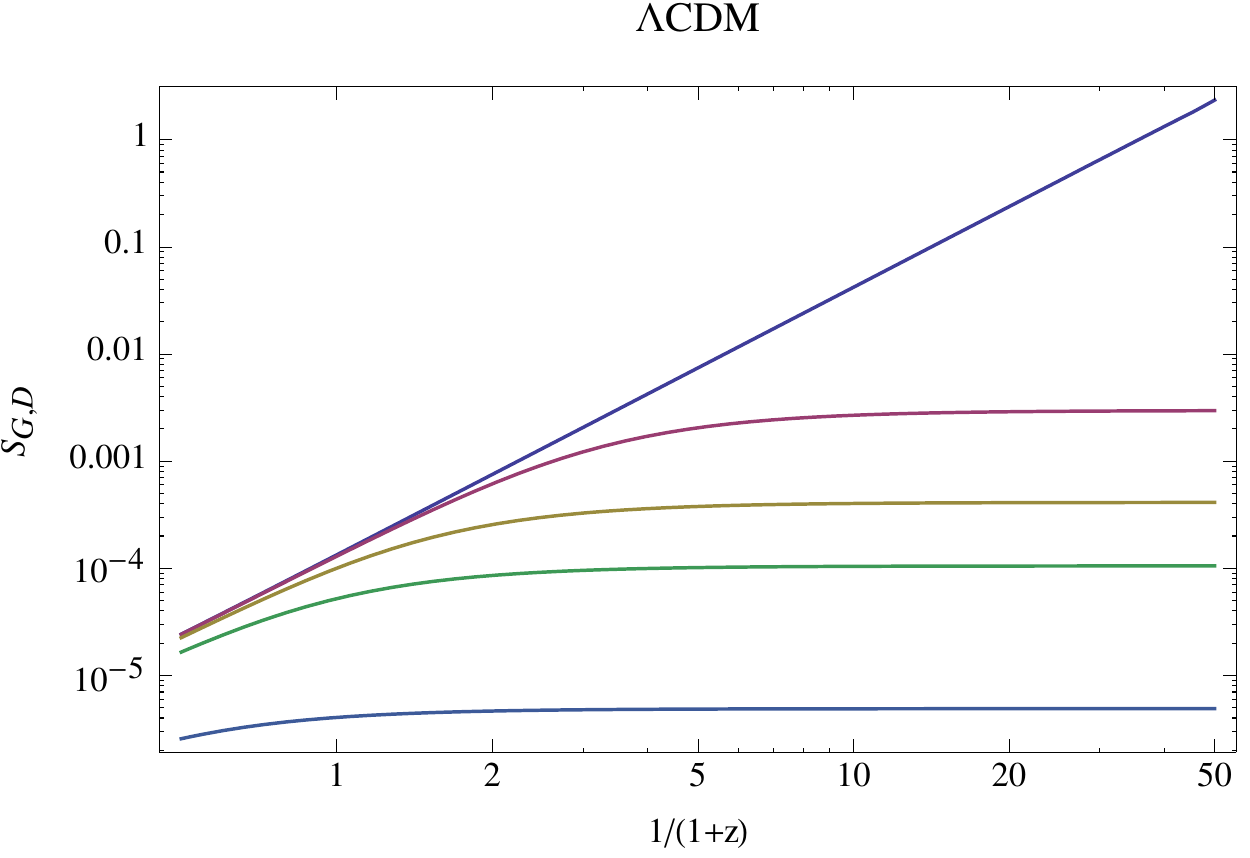}~~~~~~~~~
\centering
\caption{Value of the gravitational entropy~(\ref{DerSCliftonetall}) per unit comoving volume for a $\Lambda$CDM model and  for different values of $\Omega_{\Lambda 0}$, setting $8 \pi G=1$ and assuming an ultraviolet cut-off $k_{\rm UV}=0.1 h {\rm Mpc}^{-1}$, as a function of $1/(1+z)$. From top to bottom, $\Omega_{\Lambda 0}=0, 0.05, 0.35, 0.68$ (standard $\Lambda$CDM model), and 0.95.}
\label{Fig2}
\end{figure}

For a perturbed FL universe, at linear order, the expression~(\ref{DerSCliftonetall-second}) does not take into consideration the fundamental fact that the perturbations of our Universe are stochastic fields. To that purpose,  we need to perform both a spatial and an ensemble average. Since the Weyl scalar $C_{\mu\nu\lambda\rho}C^{\mu\nu\lambda\rho}$ is second order in perturbation, we need to first average it before we insert it  into the entropy expression. 
This corresponds to modify Eq.~(\ref{DerSCliftonetall-second}) as
\begin{equation}
\frac{S_{G,\cal D}'}{{\cal V}_{\cal D}}=4 \pi M_{Pl}^2 \lambda \frac{a}{{\cal H}}  \frac{d}{d\eta}\left(a^3 \sqrt{\frac{\overline{\langle C_{\mu\nu\lambda\rho}C^{\mu\nu\lambda\rho} \rangle_{\cal D}}}{192}}\right) 
\label{DerSCliftonetall}
\end{equation}
where $\lambda$ is a numerical constant that, similarly to what done in Ref.~\cite{CET} to recover the Bekenstein-Hawking entropy for a stationary black hole,  we fix equal to one hereafter. Using the general solution of Eq.~(\ref{Cscalarmodek}), we easily obtain the behaviour of the volume entropy for a CDM and a $\Lambda$CDM model. They are depicted on Fig.~\ref{Fig2} for several values of the cosmological constant. In all cases the gravitational entropy~(\ref{DerSCliftonetall}) is  monotonically increasing with the time. In particular, in a CDM model, the entropy goes to infinity, but with a different time behaviour as compared to the relative entropy~(\ref{Volume_Entropy_Lambda}), namely it behaves as $\sim a^{5/2}(\eta)$.

\begin{figure}[h!]
\centering
\includegraphics[width=8cm]{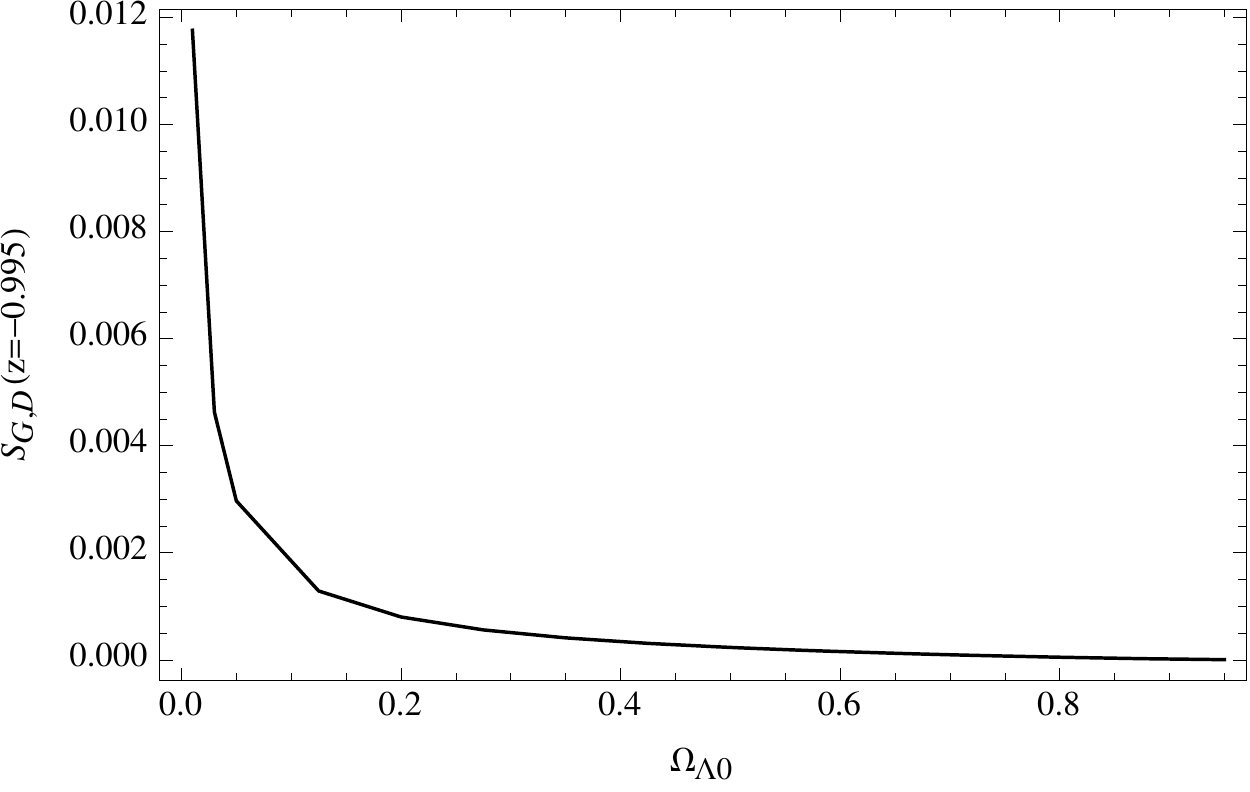}~~~~~~~~~
\centering
\caption{Asymptotic value of the gravitational entropy~(\ref{DerSCliftonetall}) per unit comoving volume for a $\Lambda$CDM universe as a function of the cosmological constant, setting $8 \pi G=1$ and assuming  an ultraviolet cut-off $k_{\rm UV}=0.1 h {\rm Mpc}^{-1}$.}
\label{Fig3}
\end{figure}

In the presence of a cosmological constant the entropy tends to a constant in the limit for which the proper time goes to infinity. 
From Fig. \ref{Fig2} one can see how the entropy asymptotically reaches its constant value, that indeed depends on the value of the cosmological constant. The maximum entropy decreases with the cosmological constant. The dependence of the maximum asymptotic value of the gravitational entropy is plotted as a function of the cosmological constant in  Fig. \ref{Fig3} and can be shown to roughly behave as $1/\Omega_{\Lambda 0}$. 

Let us try to explain the physical reasons of such a behavior for the entropy in the presence of a cosmological constant. The entropy~(\ref{DerSCliftonetall}) is the entropy associated to the formation of the large-scale structure, and is defined as an integrated effect over the cosmological history. The entropy increases when more structures are formed. When the Universe approaches the de Sitter phase, the growth of structures freezes and so does the entropy. Therefore, the gravitational entropy encodes the fact that the Universe is asymptotically de Sitter, but it only includes the entropy associated with the formation of structures. 

\begin{figure}[h!]
\centering
\includegraphics[width=8cm]{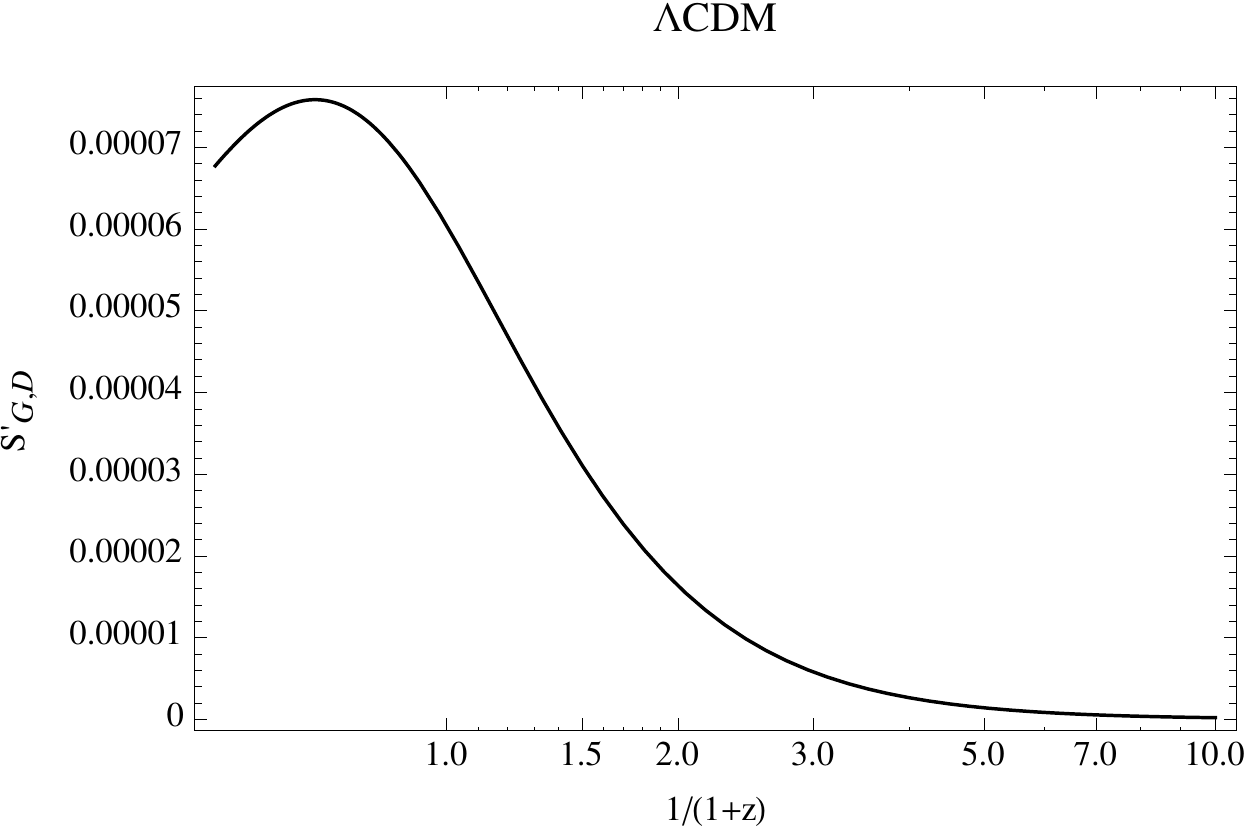}~~~~~~~~~
\centering
\caption{Value of the gravitational entropy derivative with respect to $y=1/(1+z)$ per unit comoving volume,  
obtained using  Eq.(\ref{DerSCliftonetall}), for a $\Lambda$CDM universe, setting $8 \pi G=1$ and assuming an ultraviolet cut-off $k_{\rm UV}=0.1 h {\rm Mpc}^{-1}$, in function of $y$.}
\label{Fig4}
\end{figure}

To better show this last point, we finally plot the evolution of the derivative of $S_{G,\cal D}/{\cal V}_{\cal D}$ with respect to $y=1/(1+z)$ in Fig. \ref{Fig4}. The curve has a turning point when the cosmological constant starts to dominate the expansion of the Universe and then goes to zero. As a consequence, the integrated effect stops.

\section{Discussion}\label{sec5}

In this manuscript we have computed the entropy of the large-scale structure of the Universe starting from the proposed definitions of Refs.~ \cite{CET} and~\cite{Li:2012qh}. Our results are valid in the regime for which the Universe can be described as a perturbed FL spacetime. This is valid at late times on scales $\sim10 \, h^{-1} {\rm Mpc}$ or more, where the linear regime holds. 

We have argued that the entropy should be evaluated using a spatial averaging procedure.
This is  because the entropy is not an observable in the standard way.
It describes the large-scale structure of the Universe and should arise from an average made over an extended region.

Starting from the definition of Ref.~\cite{Li:2012qh} we obtain the relative information entropy~(\ref{Volume_Entropy_Lambda}).  As shown in Fig. \ref{Fig1}, the expression~(\ref{Volume_Entropy_Lambda}) is monotonically increasing with time for a CDM model, while it does not satisfy the Penrose conjecture \cite{Penrose1,Penrose2} for a $\Lambda$CDM model. Therefore, this definition does not seem to be a valid definition of entropy for a $\Lambda$CDM universe. In fact, in the limit for which the proper time $t\rightarrow +\infty$, the importance of the cosmological constant increases and  both  FL and  perturbed FL universes can be well approximated by a de Sitter universe. Therefore, the relative information entropy stops growing at some stage and then asymptotically decreases to zero.

On the other hand, starting from the definition~\cite{CET}  we obtain the gravitational entropy~(\ref{DerSCliftonetall}). In this case,  the entropy always grows with time, both for a CDM and a $\Lambda$CDM model, hence satisfying the Penrose conjecture~\cite{Penrose1,Penrose2}. While it tends to an infinite value for a CDM model, it saturates to a constant value in the presence of a cosmological constant.  The entropy~(\ref{DerSCliftonetall}) is the entropy associated to the formation of the large scale structure of the Universe. 

The two proposals of Ref.~\cite{CET} and Ref.~\cite{Li:2012qh} were already compared in the literature in the case of a LTB dust model in Ref.~\cite{ltb2}. It was shown how in both cases the entropy of a LTB dust model grows with time satisfying the Penrose conjecture. Such models could be used only to describe the local universe and, therefore apply  in this restrict regime. Our results instead apply on large scale, where the linear and mildly non-linear regimes are valid and the Universe can be describe by a perturbed FL spacetime. The entropy obtained here can be seen as the entropy of the large-scale structure of the Universe and is complementary to the one obtained in Ref.~\cite{ltb2}.

To conclude, in \cite{Sussman:2015bea} the proposal of Ref.~\cite{CET} was studied in a non-perturbative context, with local voids of 
$50-100 \,{\rm Mpc}$ described by spherical under-dense regions with negative spatial curvature and dynamics determined by LTB dust models. 
The results obtained in \cite{Sussman:2015bea} for the gravitational entropy are in good quantitatively and qualitatively agreement with the result  presented here,
when the LTB evolution is in its linear regime. In particular, the gravitational entropy has an asymptotic behavior similar to the case of the gravitational entropy of the large-scale structure for a $\Lambda$CDM model. Therefore, the results of \cite{Sussman:2015bea} provides an important local physics connection to the large scale linear regime described here.

\acknowledgements{
We wish to thank Julien Larena for useful discussions.
GM was partially supported by the Marie Curie IEF, Project NeBRiC - ``Non-linear effects and backreaction in classical and quantum cosmology".
The work of JPU made in the ILP LABEX (under reference ANR-10-LABX-63) was supported by French state funds managed by the ANR  within the Investissements d'Avenir programme under reference ANR-11-IDEX-0004-02 and by the ANR VACOUL, ANR-10-BLAN-0510.
OU is supported by the South African Square Kilometre Array (SKA) project and CC is supported by the South African National Research Foundation (NRF).}


\appendix
\section{Dynamics of the background spacetime}\label{appA}

We assume that the background spacetime is well-described by a spatially Euclidean Friedmann-Lema\^{\i}tre universe with the late time dynamics dictated only by pressureless matter and cosmological constant with density parameters
\begin{equation}
 \Omega_{\rm m0} = \frac{8\pi G\rho_{\rm m0}}{3H_0},\qquad
 \Omega_{\Lambda0} = \frac{\Lambda}{3H_0}
\end{equation}
that satisfy $\Omega_{\rm m0}+ \Omega_{\Lambda0}=1$. The Friedmann equation then takes the usual form
\begin{equation}
 \frac{H^2(z)}{H_0^2} =   \Omega_{\rm m0}(1+z)^3 + \Omega_{\Lambda0}\,,
\label{A2}
\end{equation}
and, using the proper time $t$, its solution is given by
\begin{equation}
 a(t) \propto \sinh^{2/3}\left(\frac{3}{2}\sqrt{\Omega_{\Lambda0}} H_0t \right).
\end{equation}
The normalization to the Hubble constant today, $H_0$, implies that
\begin{equation}
  \sinh\left(\frac{3}{2}\sqrt{\Omega_{\Lambda0}} H_0t_0 \right) = \frac{\Omega_{\Lambda0}^{1/2}}{(1-\Omega_{\Lambda0})^{1/2}}\equiv \kappa_0^{3/2}
\end{equation}
so that the redshift is given by
\begin{equation}
 1+z =\frac{\kappa_0}{\sinh^{2/3}\left(\frac{3}{2}\sqrt{\Omega_{\Lambda0}} H_0t \right)}.
\end{equation}


\section{Linear perturbation theory}\label{appB}

The evolution of the degrees of freedom of the metric~(\ref{General_metric}) can be found in many textbooks, e.g. Ref.~\cite{pubook}. 

It can first be shown that there is only 6 gauge invariant degrees of freedom usually defined as
\begin{eqnarray}
\Psi &\equiv& \psi +\frac{1}{2} \Delta E+\frac{\mathcal{H}}{2}(\beta+E')\,, \label{b1}\\
\Phi &\equiv& \alpha-\frac{\mathcal{H}}{2}(\beta+E')-\frac{1}{2}(\beta+E')'\,, \\
\bar{\Phi}^i &\equiv& \frac{1}{2}\bar{\chi}^{i'}+\frac{1}{2}\bar{B}^{i}\,\,\,\,\,\,\,\,,\,\,\,\,\,\,\,
\bar{h}^{i j} \,,
\label{b4}
\end{eqnarray}
where the prime denotes the derivative with respect to conformal time and ${\cal H}=a'/a$.

Considering a matter sector described by a perfect fluid with stress-energy tensor
\begin{equation}
T_{\mu \nu}=(\rho+P)u_\mu u_\nu+P g_{\mu\nu}\,,
\end{equation}
where the density and pressure can be split as $\rho(\eta, {\bm x})=\rho^{(0)}(\eta)+\rho^{(1)} (\eta, {\bm x})$ and 
$P(\eta, {\bm x})=P^{(0)}(\eta)+P^{(1)} (\eta, {\bm x})$, and the velocity of the comoving observers is decomposed as $u^\mu=\bar{u}^\mu+\delta u^\mu$ with $u_\mu u^\mu=-1$. It follows that
\begin{equation}
u^\mu=a^{-1}(1-\alpha, v^i)\,\,\,\,,\,\,\,u_\mu=a (-1-\alpha, v_i-1/2  B_i)\,
\end{equation}
and we decompose $v_i$ into scalar and a vector component according to
\begin{equation}
  v_i=\partial_i v+\bar{v}_i\,.
\end{equation}

The scalar shear, given in Eq.~(\ref{sigma2}), is by construction a second order quantity so that
\begin{equation} \label{shear0and1}
 (\sigma^2)^{(0)}= (\sigma^2)^{(1)}=0.
\end{equation}
Thus, at the lowest order and for a general metric, we have
\begin{eqnarray}
(\sigma^2)^{(2)}&=&\frac{1}{2 a^2 {S'}^{2}}\left[\delta S_{,i j}
\delta S^{,i j}-\frac{1}{3}(\nabla^2 \delta S)^2 \right]
\nonumber \\ & &
+\frac{1}{8 a^2}\left[B_{i, j}
B^{i, j}-\frac{1}{3}(\partial^i B_i)^2 \right] \nonumber\\
& & -\frac{1}{2 a^2 S'}\left[\delta S_{,i j}
B^{i, j}-\frac{1}{3}(\nabla^2 \delta S) (\partial^i B_i) \right]
\nonumber \\ & &
-\frac{1}{a^2 {S'}} \delta S_{,i j} {\tilde{h}}^{',i
j}+\frac{1}{2 a^2}  B_{i, j} {\tilde{h}}^{',i j}
\nonumber \\ & &
+\frac{1}{2 a^2}
{\tilde{h}}_{',i j} {\tilde{h}}^{',i j}\,,
\label{sigma2order2general}
\end{eqnarray}
where  $\tilde{h}_{,i j}=\frac{1}{2}D_{ij} E+\partial_{(i} \bar{\chi}_{j)}+\frac{1}{4}\bar{h}_{i j}$, we use the notation $X_{,i}\equiv \partial_i X$ for any field $X$, and $\delta S$ is the first order perturbation of the scalar $S({\bf x}, t)$ defining the space-time foliation.  

It is clear that first order perturbation theory is sufficient to obtain the general expression for the shear up to second order, since second order perturbations will contribute only to third or fourth order to $\sigma^2$.


\end{document}